\documentclass[%
reprint,
superscriptaddress,
 amsmath,amssymb,
 aps,
 prl,
floatfix,
]{revtex4-1}

\usepackage{graphicx}% Include figure files
\usepackage{dcolumn}% Align table columns on decimal point
\usepackage{bm}% bold math
\usepackage{siunitx}
\usepackage{color}

\begin{document}

\preprint{}

\title{Nonlinear longitudinal and transverse magnetoresistances due to current-induced magnon creation-annihilation processes}% Force line breaks with \\

\author{Paul Noël}
\email{paul.noel@ipcms.unistra.fr\\
Present address: Université de Strasbourg, CNRS, Institut de Physique et Chimie des Matériaux de Strasbourg, UMR 7504, Strasbourg F-67000, France}
\affiliation{Department of Materials, ETH Zurich, CH-8093 Zurich, Switzerland}
\author{Richard Schlitz}
\affiliation{Department of Materials, ETH Zurich, CH-8093 Zurich, Switzerland}
\author{Emir Karadža}
\affiliation{Department of Materials, ETH Zurich, CH-8093 Zurich, Switzerland}
\author{Charles-Henri Lambert}
\affiliation{Department of Materials, ETH Zurich, CH-8093 Zurich, Switzerland}
\author{Luca Nessi}
\affiliation{Department of Materials, ETH Zurich, CH-8093 Zurich, Switzerland}
\affiliation{Dipartimento di Fisica, Politecnico di Milano, Via G. Colombo 81, 20133 Milano, Italy}
\author{Federico Binda}
\affiliation{Department of Materials, ETH Zurich, CH-8093 Zurich, Switzerland}
\author{Pietro Gambardella}
\email{pietro.gambardella@mat.ethz.ch\\
}
\affiliation{Department of Materials, ETH Zurich, CH-8093 Zurich, Switzerland}

\date{\today}% It is always \today, today,

\begin{abstract}

Charge-spin conversion phenomena such as the spin Hall effect allow for the excitation of magnons in a magnetic layer by passing an electric current in an adjacent nonmagnetic conductor. We demonstrate that this current-induced modification of the magnon density generates an additional nonlinear longitudinal and transverse magnetoresistance for every magnetoresistance that depends on the magnetization. Using harmonic measurements, we evidence that these magnon creation-annihilation magnetoresistances dominate the second harmonic longitudinal and transverse resistance of thin Y$_{3}$Fe$_{5}$O$_{12}$/Pt bilayers. Our results apply to both insulating and metallic magnetic layers, elucidating the dependence of the magnetoresistance on applied current and magnetic field for a broad variety of systems excited by spin currents.

\end{abstract}

%Use showkeys class option if keyword
\maketitle

%%%%%%%%%%%%%%%%%%%%%%%%%%%%%%%% INTRODUCTION %%%%%%%%%%%%%%%%%%%%%%%%%%%%%%%%%%

Magnons are known to influence electric transport in ferromagnets (FM)~\cite{kasuya56}. The scattering of electrons with thermally excited magnons induces a negative magnon magnetoresistance (MMR), as the magnon population decreases under an applied magnetic field~\cite{raquet02, mihai08}. Additional effects emerge when a FM is coupled with a nonmagnetic metal (NM) forming a FM/NM bilayer. In such systems, an electric current flowing in the NM leads to a spin accumulation at the FM/NM interface via charge-to-spin conversion processes promoted by spin-orbit coupling, such as the spin Hall effect~\cite{sinova15,amin16}. The spin accumulation, in turn, modulates the magnon population in the FM~\cite{demidov11, demidov17} and can be used to compensate the intrinsic damping of the magnetic layer~\cite{ando08}. This mechanism is actively exploited in spin Hall nano-oscillators, holding potential as an energy-efficient source of coherent magnons~\cite{collet16}. The reverse effect, i.e., the electrical detection of magnons, is routinely employed as a sensitive probe of thermally-induced magnon currents~\cite{uchida08}, long range spin transport~\cite{chumak12,cornelissen15, goennenwein15}, and spin pumping by coherent microwave absorption~\cite{tserkovnyak05, saitoh06}. 

\begin{figure}[ht]
\includegraphics[width=\columnwidth]{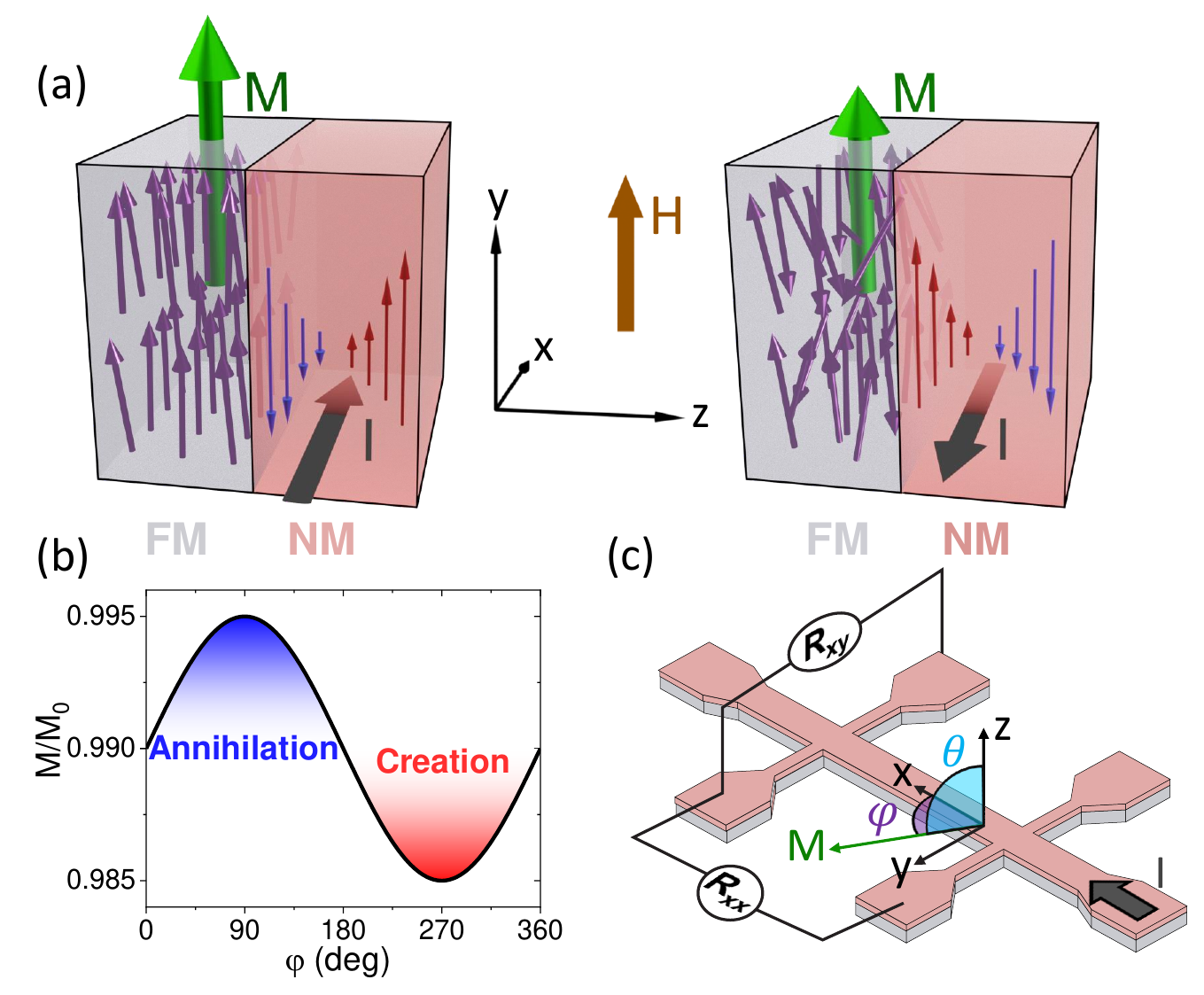}
\caption{\label{fig:figure1} (a) Schematic of the annihilation and creation of magnons in a FM due to current-induced spin accumulation in the NM. The local magnetic moments are shown as purple arrows, the spin moments are shown as thin red and blue arrows. The spin magnetic moment is opposite to the spin moment. (b) Expected change of the magnetization as a function of the angle $\varphi$. (c) Schematic of the sample and coordinate system.}
\end{figure}

The excitation of magnons by an interfacial spin accumulation depends on the relative direction of the accumulated spins and the magnetization~\cite{demidov11, cornelissen15}. Spin-flip scattering leads to the creation (annihilation) of magnons when the magnetization $M$ is parallel (antiparallel) to the spin accumulation, as depicted in Fig.~\ref{fig:figure1}(a). In turn, the modification of the magnon population by a spin current leads to a change of $M$, as depicted in Fig.~\ref{fig:figure1}(b). Because only the component of the spin accumulation collinear to $M$ contributes to magnon creation-annihilation, the magnetization as a function of current $I$ is described by
\begin{equation}
M (I) = M_{s} + \Delta M (I) \sin\varphi,
\label{eq:one}
\end{equation}
where $M_{s}$ is the saturation magnetization, $\Delta M(I)$ is the variation of $M$ due to current-induced changes of the magnon population and $\varphi$ the in-plane angle between $M$ and $I$. Due to thermally-excited magnons, $M_{s}$ is always smaller than the spontaneous magnetization at zero temperature $M_{0}$.  

In FM/NM bilayers, changes in magnetization due to variations in the magnon population give rise to nonlinear magnetoresistances. Whereas nonlocal magnetotransport effects driven by magnons are well established~\cite{cornelissen15, goennenwein15, thiery18a, gao22}, the influence of magnon creation-annihilation on local electron transport has received comparatively less attention, except for the unidirectional spin Hall magnetoresistance (USMR)~\cite{avci15, langenfeld16, yasuda16, li17, wang18, borisenko18, avci18,sterk19, liu21a, chang21, chen22}. However, as we argue below, these effects are general to \textit{all} manifestations of magnetoresistance in magnetic layers in the presence of electrically-induced spinand orbital currents, \color{black} and should be taken into account for a correct description of magnetotransport phenomena.These include the different types of longitudinal and transverse magnetoresistances commonly found in magnetic conductors as well as in FM/NM bilayers with insulating FM. Additionally, we find that the magnon-induced component of the USMR identified in prior work~\cite{avci15, langenfeld16, yasuda16, li17, wang18, borisenko18, avci18, sterk19, kim19, liu21a, chang21, chen22} must be corrected to properly reflect its dependence on magnetization.
\color{black}

In this letter, we provide a complete description of the magnon-induced nonlinear longitudinal and transverse magnetoresistances, and study their manifestation in Y$_{3}$Fe$_{5}$O$_{12}$(YIG)/Pt  bilayers. We term this set of nonlinear effects the magnon creation-annihilation magnetoresistances (\emph{m}$\!^{\dagger}$\!\emph{m}MRs). The \emph{m}$\!^{\dagger}$\!\emph{m}MRs are distinct phenomena but have angular dependencies similar to other nonlinear magnetoresistances that presume a constant magnetization and originate from spin-orbit torques (SOT)~\cite{avci14}, spin-dependent USMR~\cite{avci18} and magnetothermal effects like the anomalous Nernst effect (ANE)~\cite{miyasato07} and spin Seebeck effect (SSE)~\cite{uchida08}. Further, we show that the \emph{m}$\!^{\dagger}$\!\emph{m}MRs can be used to probe the current-induced magnon population and damping compensation in FM/NM bilayers. In a joint paper, we demonstrate the presence of strong \emph{m}$\!^{\dagger}$\!\emph{m}MRs in metallic FM/NM systems and evidence how they affect the electrical detection of SOTs in both conducting and insulating magnets~\cite{jointpaper}.  \\
Changes of the magnon population are known to affect the magnetization and hence the amplitude of the magnetoresistance. The anisotropic magnetoresistance (AMR) has been studied in alloys with low Curie temperature and found to vary with $M^2$~\cite{hamzic78,baxter02}, in line with predictions~\cite{parker51}. Similarly, the spin Hall magnetoresistance (SMR) is proportional to $M^2$~\cite{nakayama13,uchida15, lammel19, zhang19}. As a consequence, the planar Hall effect (PHE) resulting from either the AMR~\cite{ky68} or SMR~\cite{chen13} also depends on $M^2$, whereas the anomalous Hall effect (AHE) scales linearly with the out-of-plane magnetization $M_z$~\cite{nagaosa10}. The AMR, SMR, and PHE also depend on the direction of the magnetization relative to the current. 
Accounting for these dependencies, the following proportionality relationships hold: $R_\mathrm{AMR} \propto M_{x}^{2}$, $R_\mathrm{SMR} \propto - M_{y}^{2}$, $R_\mathrm{PHE} \propto M_{x}M_{y}$, and $R_\mathrm{AHE} \propto M_{z}$, where $M_{x}$ and $M_{y}$ are the magnetization in the sample plane along and perpendicular to the current, respectively [see Fig.~\ref{fig:figure1} (c)]. The MMR associated with electron-magnon scattering is sensitive only to the total magnon population and the effective field ~\cite{gil05}. Using Eq.~(\ref{eq:one}) and disregarding the terms proportional to $\Delta M(I)^{2}$ we obtain the following expressions for the magnetoresistances: \begin{eqnarray}
R_\mathrm{AMR} \propto M_{s}^{2}\cos^{2}\varphi+2M_{s}\Delta M(I) \sin\varphi\cos^{2}\varphi,
\label{eq:a}\\
R_\mathrm{SMR} \propto  -M_{s}^{2}\sin^{2}\varphi-2M_{s}\Delta M(I) \sin^{3}\varphi,
\label{eq:b}\\
R_\mathrm{PHE} \propto  M_{s}^{2}\sin\varphi\cos\varphi+2M_{s}\Delta M(I) \sin^{2}\varphi\cos\varphi,
\label{eq:c}\\
R_\mathrm{AHE} \propto  M_{s}\cos(\theta)+\Delta M(I) \sin\varphi\cos\theta,
\label{eq:d}\\
R_\mathrm{MMR} \propto [M_{0}-M_{s}]+\Delta M(I)\sin\varphi,
\label{eq:e}
\end{eqnarray}
where, for brevity, the out-of-plane angle $\theta$ is only explicitly included for the AHE. The first term on the right-hand side represents the usual angular dependence of each magnetoresistance and is current independent, whereas the second term describes the \emph{m}$\!^{\dagger}$\!\emph{m}MR contributions. It is possible to separate the two terms by performing harmonic resistance measurements using an alternate current at the frequency $\omega/2\pi$. The current-independent magnetoresistances $R_\mathrm{AMR}^{1\omega}$, $R_\mathrm{SMR}^{1\omega}$, $R_\mathrm{PHE}^{1\omega}$, $R_\mathrm{AHE}^{1\omega}$ and $R_\mathrm{MMR}^{1\omega}$ are then detected as a first harmonic signal. The second term, proportional to the spin current-induced change of the magnon population, gives rise to the nonlinear \emph{m}$\!^{\dagger}$\!\emph{m}MRs that can be detected as a second harmonic signal:
\begin{eqnarray}
R_\mathrm{AMR}^{2\omega}= 2R_\mathrm{AMR}^{1\omega}\frac{\Delta M(I)}{M_{s}}\sin\varphi\cos^{2}\varphi,
\label{eq:f}
\\
R_\mathrm{SMR}^{2\omega}=-2R_\mathrm{SMR}^{1\omega}\frac{\Delta M(I)}{M_{s}}\sin^{3}\varphi,
\label{eq:g}
\\
R_\mathrm{PHE}^{2\omega} = 2R_\mathrm{PHE}^{1\omega}\frac{\Delta M(I)}{M_{s}}\sin^{2}\varphi\cos\varphi,
\label{eq:h}
\\
R_\mathrm{AHE}^{2\omega} = R_\mathrm{AHE}^{1\omega}\frac{\Delta M(I)}{M_{s}}\sin\varphi\cos\theta,
\label{eq:i}
\\
R_\mathrm{MMR}^{2\omega} \propto \Delta M(I)\sin\varphi.
\label{eq:j}
\end{eqnarray}
For the sake of simplicity we limit our description to the most common longitudinal and transverse magnetoresistances observed in FM/NM bilayers for a spin accumulation along $y$ and rotations of the magnetization in the sample plane, and the AHE. Other \emph{m}$\!^{\dagger}$\!\emph{m}MRs can be derived similarly. Also, we disregard higher-order effects in Eq.~(\ref{eq:one}) that occur close to the damping compensation~\cite{thiery18a, wimmer19, guckelhorn21}, which would give rise to higher harmonic terms. 
 We note that several effects contribute to the magnonic USMR, namely the AMR (Eq.~\ref{eq:f}), SMR (Eq.~\ref{eq:g}), and MMR (Eq.~\ref{eq:j}), when $\Delta M$ is induced by the absorption of a spin current. Thus, the angular dependence of the USMR is $\sin^{3}\varphi + \sin\varphi$, in contrast to the $\sin\varphi$ dependence assumed in prior studies~\cite{li17, avci18, liu21a, chang21, chen22}. Previous models accounted only for the electron scattering asymmetry due to variations in magnon population, overlooking the associated change in magnetization intensity. \color{black} 

The amplitude of the \emph{m}$\!^{\dagger}$\!\emph{m}MRs is more pronounced for large changes in the magnon population, which are easily obtained in magnetic thin films with low damping and low $M_{s}$. Hence, we performed harmonic magnetoresistance measurements in a YIG/Pt bilayer with ultra-thin YIG (6.2~nm). YIG/Pt is  extensively studied for its magnonic properties, and the damping compensation in thin films of YIG is obtained for current densities $J$ of a few $10^{11}$ A/m$^{2}$~\cite{cornelissen15,collet16, wimmer19, guckelhorn21}. A large modification of the magnon population is therefore easily accessible. The magnetoresistive response of YIG/Pt is also well known, with longitudinal SMR and associated PHE and the absence of competing magnetoresistive effects due to AMR or MMR~\cite{chen13, althammer13, nakayama13}. 

The YIG film was grown by pulsed laser deposition on a (111)-oriented Gd$_{3}$Ga$_{5}$O$_{12}$ substrate. A $3$~nm Pt film was deposited by sputtering in situ to preserve a good interface quality~\cite{mendil19a, mendil19b}. The sample was subsequently patterned into Hall bars of width $w = 10~\mu$m and length $L=50~\mu$m, with an aspect ratio $g=L/w \approx 5$. We probed the in-plane angular dependence of the first and second harmonic longitudinal resistance ($R_{xx}^{1\omega,2\omega}$) and transverse resistance ($R_{xy}^{1\omega,2\omega}$) of YIG/Pt as a function of magnetic field $H$. The measurements were carried out at room temperature using an alternate current of amplitude $I = 4$~mA ($J = 1.33 \times 10^{11}$~A/m$^{2}$) at a frequency $\omega/(2\pi) = \SI{10}{\hertz}$. The absence of AMR in our samples was confirmed by magnetoresistance measurements~\cite{supmat}. 

\begin{figure}[ht!]
\includegraphics[width=\columnwidth]{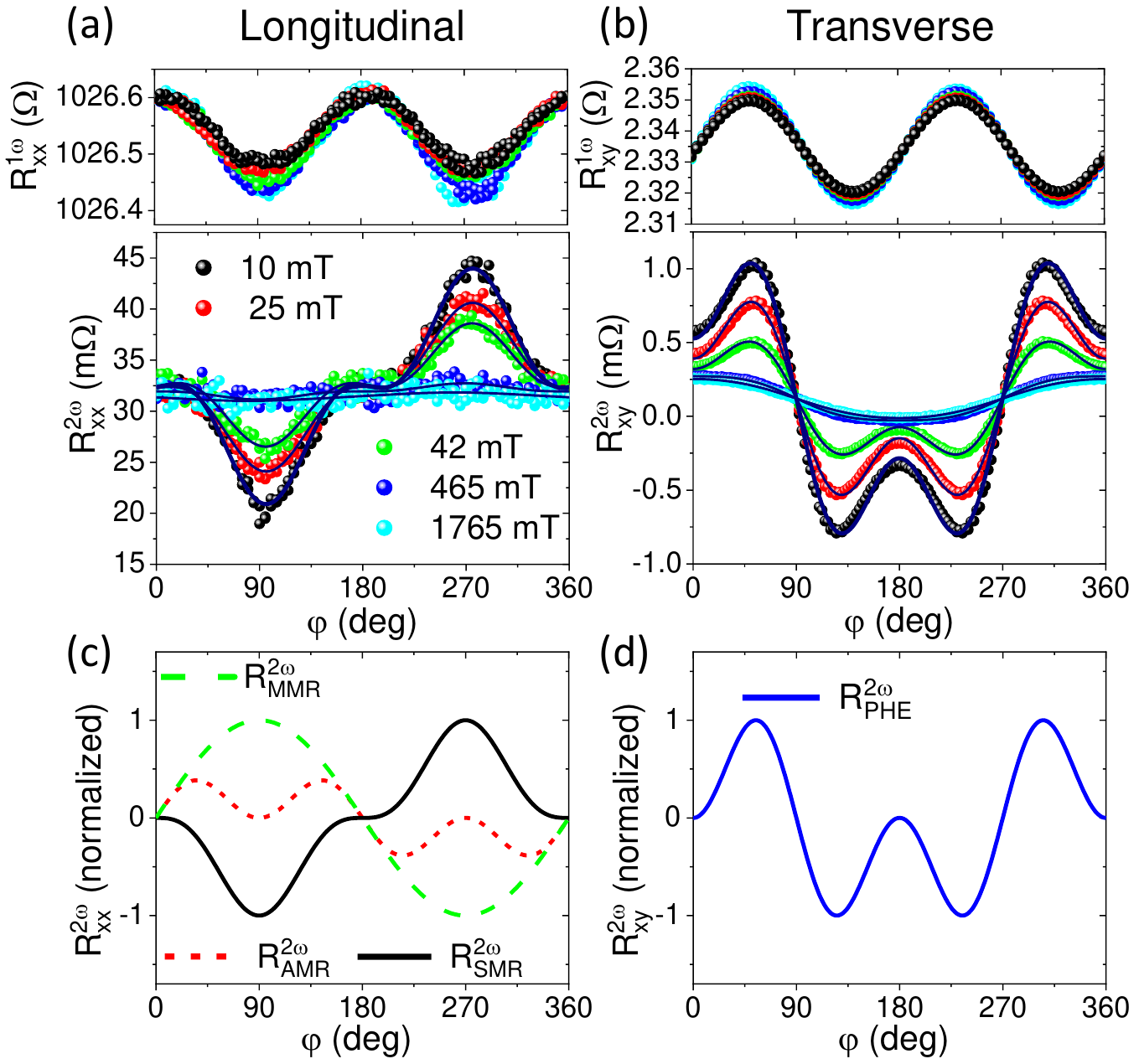}
\caption{\label{fig:figure2}  Angular dependence of the first and second harmonic (a) longitudinal and (b) transverse resistance of YIG/Pt for different values of the applied field at a current of $4$ mA. The solid lines are fits to Eqs.~(\ref{eq:k}) and (\ref{eq:l}). (c, d) Expected angular dependence of the longitudinal and transverse second harmonic resistance due to the magnon creation annihilation according to Eqs.~(\ref{eq:f}-\ref{eq:j}).}
\end{figure}

Figure~\ref{fig:figure2} (a) and (b) show $R_{xx}^{1\omega,2\omega}$ and $R_{xy}^{1\omega,2\omega}$ as a function of $\varphi$ for different values of the applied magnetic field $H$. $R_{xx}^{1\omega}$ and $R_{xy}^{1\omega}$ follow the typical angular dependence of the SMR and PHE, respectively, as expected from Eqs.~(\ref{eq:b}) and (\ref{eq:c}). $R_{xx}^{2\omega}$ follows a $\sin^{3}\varphi$ dependence, consistent with the nonlinear SMR contribution due to magnon creation annihilation, i.e., the \emph{m}$\!^{\dagger}$\!\emph{m}SMR [Eq.~(\ref{eq:g})]. $R_{xy}^{2\omega}$ shows a prominent $\sin^{2}\varphi\cos\varphi$ dependence, consistent with a dominant nonlinear PHE contribution due to magnon creation annihilation, i.e., the\emph{m}$\!^{\dagger}$\!\emph{m}PHE  [Eq.~(\ref{eq:h})]. These nonlinear signals are quickly suppressed upon increasing the magnetic field, as expected for magnonic effects~\cite{thiery18a, avci18, borisenko18}. Figure \ref{fig:figure2} (c) and (d) summarize the in-plane angular dependence of the different longitudinal and transverse \emph{m}$\!^{\dagger}$\!\emph{m}MR contributions to the 2nd harmonic resistance according to Eqs.~(\ref{eq:f}-\ref{eq:j}). 
\begin{figure}[ht]
\includegraphics[width=\columnwidth]{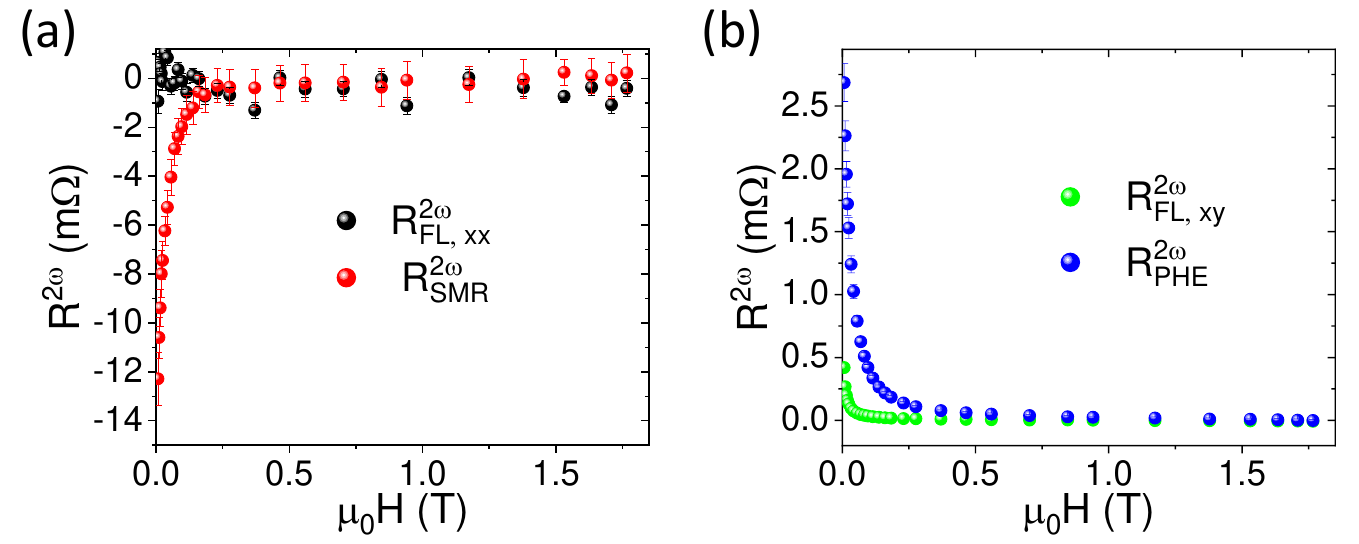}
\caption{\label{fig:figure3} Magnetic field dependence of (a) the second harmonic SMR and FL-SOT contributions to the longitudinal resistance and (b) the second harmonic PHE and FL-SOT contributions to the transverse resistance.
The data points are obtained from the fits of $R_{xx}^{2\omega}$ and $R_{xy}^{2\omega}$ using Eqs.~\ref{eq:k} and \ref{eq:l}.
}
\end{figure}

In principle, additional effects contribute to $R_{xx}^{2\omega}$ and $R_{xy}^{2\omega}$, such as the SOT-induced oscillations of $M$~\cite{avci14}, the spin-dependent USMR~\cite{avci15, avci18}, and magnetothermal effects~\cite{miyasato07, uchida08}. Differentiating the \emph{m}$\!^{\dagger}$\!\emph{m}MRs contributions from the SOT and magnetothermal contributions is not straightforward due to similarities in their angular dependence (see Ref.~\onlinecite{jointpaper} for a more detailed discussion). In this regard YIG/Pt is an excellent model system, because the AMR, MMR, ANE and spin-dependent USMR \color{black} are absent owing to the insulating nature of YIG. Additionally,  due to the small AHE, the dampinglike (DL) SOT contribution $R_{\mathrm{DL}, xy}^{2\omega}$ is negligible~\cite{supmat}. The longitudinal and transverse SSE contributions ($R_{\mathrm{SSE}, xx}^{2\omega}$ and $R_{\mathrm{SSE}, xy}^{2\omega}$) are field-independent and can be extracted at the highest measurement field when other contributions are negligible. \color{black} The angular dependence of the second harmonic resistance accounting for all the expected contributions can thus be written as
\begin{eqnarray}
    R_{xx}^{2\omega}= R_{\sin}^{2\omega}\sin\varphi + R_{\sin^{3}}^{2\omega}\sin^{3}\varphi,
        \label{eq:k} \\   
    R_{xy}^{2\omega}=R_{\cos}^{2\omega}\cos\varphi + R_{\cos^{3}}^{2\omega}\cos^{3}\varphi,
        \label{eq:l}
\end{eqnarray}
where $R_{\cos}^{2\omega}=R_\mathrm{PHE}^{2\omega}-R_{\mathrm{FL}, xy}^{2\omega}+R_{\mathrm{SSE}, xy}^{2\omega}$, $R_{\cos^{3}}^{2\omega}=2R_{\mathrm{FL}, xy}^{2\omega}-R_\mathrm{PHE}^{2\omega}$, $R_{\sin}^{2\omega}=R_{\mathrm{SSE}, xx}^{2\omega}+R_{\mathrm{FL}, xx}^{2\omega}$ and $R_{\sin^{3}}^{2\omega}=R_\mathrm{SMR}^{2\omega}-R_{\mathrm{FL}, xx}^{2\omega}$. 
We fit $R_{xx}^{2\omega}$ and $R_{xy}^{2\omega}$ shown in Fig.~ \ref{fig:figure2} using Eqs.~(\ref{eq:k}) and (\ref{eq:l}) to extract the different contributions~\cite{supmat}. The results are shown in Fig.~\ref{fig:figure3}. The \emph{m}$\!^{\dagger}$\!\emph{m}PHE and \emph{m}$\!^{\dagger}$\!\emph{m}SMR contributions are strongly suppressed with magnetic field. Moreover, we find that the two effects scale with the aspect ratio of the Hall bar, $R_\mathrm{SMR}^{2\omega} = -g \cdot R_\mathrm{PHE}^{2\omega}$~\cite{supmat}, as expected because the PHE in YIG/Pt is the transverse manifestation of the SMR~\cite{chen13}. Only a small part of the second harmonic signal originates from the FL torque. From $R_{\mathrm{FL}, xy}^{2\omega}$ we estimate the FL effective field to be $0.218 \pm 0.004 $~mT ~\cite{supmat}. This is very similar to the Oersted field $B_\mathrm{Oe} = \frac{\mu_{0}I}{2w} = 0.25 \pm 0.03 $~mT calculated by assuming $I=4$~mA and an error on $w$ of up to 1 $\mu$m due to the limited resolution of optical lithography. The FL torque scales linearly with $J$ and is mostly given by the Oersted field~\cite{supmat}, in line with previous observations in thicker YIG/Pt~\cite{fang17, mendil19b}. Not accounting for the \emph{m}$\!^{\dagger}$\!\emph{m}PHE contribution would have resulted in a different sign and much higher estimate of the FL torque~\cite{jointpaper}. 

\begin{figure}[t]
\includegraphics[width=\columnwidth]{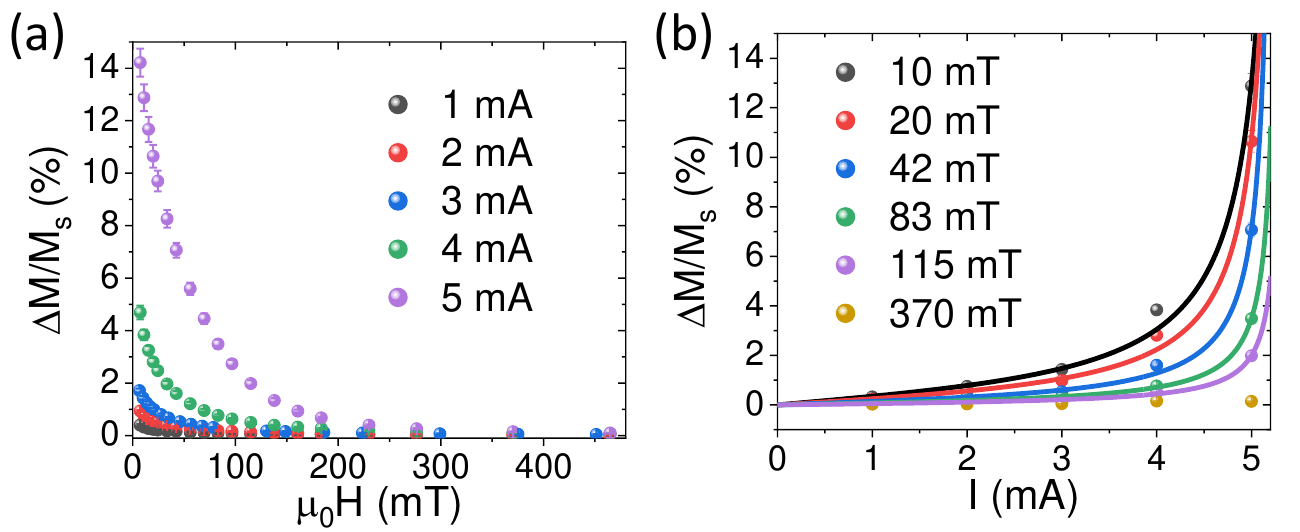}
\caption{\label{fig:figure4} Relative change of the magnetization due to current-induced magnon creation. (a) Field dependence of $\Delta M(I)/M_{s}$ for different currents. (b) Current-dependence of $\Delta M(I)/M_{s}$ for various fields. The magnetic field is applied along $y$. The solid lines are fits to the data according to Eq.~\ref{eq:nine}.}
\end{figure}

Besides providing a proper description of the current-induced magnetoresistance in FM/NM bilayers, the \emph{m}$\!^{\dagger}$\!\emph{m}MRs also allow for quantifying the relative change of the magnetization due to modifications of the magnon population, namely $\Delta M(I)$. This quantity is relevant for estimating the efficiency of magnon excitation processes and measuring the magnon population, which is usually accessible only by Brillouin light scattering and nonlocal magnetoresistance probes~\cite{demidov11,demidov17, borisenko18, thiery18a}. The \emph{m}$\!^{\dagger}$\!\emph{m}MRs provide a simple electrical probe of $\Delta M(I)$ that is readily accessible in two-terminal devices. Following Eq.~(\ref{eq:h}), the relative change of magnetization is $\frac{\Delta M(I)}{M_{s}} = \frac{R_\mathrm{PHE}^{2\omega} }{ 2R_\mathrm{PHE}^{1\omega}}$. Figure~\ref{fig:figure4}(a) shows $\Delta M(I)$ for different currents as a function of the magnetic field applied parallel to $y$. The change of the magnetization is larger for low fields and large currents and reaches up to 14 \% at 7~mT and $I=5$~mA ($J = 1.66 \times 10^{11}$~A/m$^{2}$). The reduced change of magnetization at high magnetic field indicates the dominant role of low-energy magnons, \color{black}as their relaxation is more prominently affected by the external field~\cite{collet16, thiery18a, kohno23}. In Ref.~\onlinecite{supmat} we report temperature-dependent measurements that evidence a strong decrease of $\Delta M(I)$ associated with the reduced contribution of the nonlinear effects at low temperature, while the FL torque remains constant. The suppression of the \emph{m}$\!^{\dagger}$\!\emph{m}MRs at low temperature further confirms their magnonic nature~\cite{goennenwein15, avci18}.

Finally, we find that $\Delta M(I)/M_{s}$ becomes nonlinear at large currents [Fig.~\ref{fig:figure4}(b)], in line with results obtained by Brillouin light scattering and nonlocal resistance measurements~\cite{thiery18a, wimmer19, guckelhorn21}. The current dependence of the magnetization change can be described by~\cite{borisenko18}
\begin{equation}
    \frac{\Delta M(I)}{M_{s}} \propto \frac{I/I_{c}}{1-(I/I_{c})^{2}},
    \label{eq:nine}
\end{equation}
where $I_{c}$ is the critical current at which the damping is fully compensated. Thus, measuring $\Delta M(I)$ using the \emph{m}$\!^{\dagger}$\!\emph{m}MRs offers a simple way to determine the damping compensation threshold. Equation~(\ref{eq:nine}) fits the data accurately in a broad range of fields [solid lines in Fig.~\ref{fig:figure4}(b)] and yields the critical current $I_{c}$, which is constant around $5.3 \pm 0.1$ mA below 100~mT and increases linearly for higher magnetic fields~\cite{supmat}, in agreement with previous estimates by nonlocal measurements~\cite{wimmer19, guckelhorn21}. We observed a similar current dependence in the \emph{m}$\!^{\dagger}$\!\emph{m}SMR detected as a change in the longitudinal resistance~\cite{supmat}. For currents above 5~mA the resistivity of the Pt layer was permanently modified and the quality of the interface was reduced due to Joule heating~\cite{schlitz21}. Therefore, we could not reach beyond $I_{c}$ in these structures. \\

In conclusion, we provided evidence of a general class of nonlinear magnetoresistances that are due to the excitation of magnons caused by the injection of a spin current in FM/NM bilayers. These \emph{m}$\!^{\dagger}$\!\emph{m}MRs exist in FM/NM systems with either conductive or insulating FM and involve every type of longitudinal or transverse magnetoresistance that depends on the magnetization. 
Our results extend beyond the magnonic contribution to the USMR reported previously~\cite{langenfeld16, li17, wang18, borisenko18,avci18, sterk19, liu21a} and show that the current-induced creation and annihilation of magnons should be taken into account to achieve a consistent description of all magnetoresistive phenomena with both even and odd dependence on the magnetization. 

The nonlinear magnetoresistances that we describe do not require any spin momentum locking or hexagonal warping of the band structure and thus differ from the ones obtained in magnetic topological insulators~\cite{yasuda16, yasuda17}, topological materials~\cite{he18} and Rashba interfaces~\cite{guillet20}. In particular, the nonlinear PHE described in this work originates from magnons in contrast to the nonlinear PHE observed in topological insulators and Rashba interfaces~\cite{he19} or the nonlinear Hall effect due to the Berry curvature dipole~\cite{du21}.

Our results evidence that the \emph{m}$\!^{\dagger}$\!\emph{m}MRs appear prominently in the second harmonic longitudinal and transverse magnetoresistances of thin ferrimagnetic YIG/Pt bilayers. As shown in Ref.~\onlinecite{jointpaper}, these effects also exist in conductive FM/NM bilayers but were previously confused with other nonlinear effects due to SOTs and magnetothermal effects. The \emph{m}$\!^{\dagger}$\!\emph{m}MRs can strongly affect the estimation of the SOTs using harmonic Hall measurements. Moreover, the \emph{m}$\!^{\dagger}$\!\emph{m}MRs enable the simple all-electrical detection of current-induced changes of the local magnon population. This opens the possibility for a local electrical detection of the damping compensation and consequently of magnon Bose-Einstein condensates~\cite{borisenko20, schneider21}. 

 The \emph{m}$\!^{\dagger}$\!\emph{m}MRs are additionally anticipated in materials with reduced symmetry and unconventional spin accumulation with both $x$ and $y$ components~\cite{macneill17,roy22}, as shown in Ref.~\cite{supmat}. \color{black} Due to the universality of the effect, the \emph{m}$\!^{\dagger}$\!\emph{m}MRs are also expected to appear in antiferromagnet/NM bilayers that exhibit PHE and SMR~\cite{hoogeboom17}, where the change of the magnon population would affect the Néel vector instead of the magnetization, and also in systems that exhibit orbital magnetoresistances~\cite{okano19,ding22a,ding22b}. 

\begin{acknowledgments}
We acknowledge Morgan Trassin for providing the YIG/Pt sample. Discussions with Sa\"ul V\'elez are gratefully acknowledged. This work was supported by the Swiss National Science Foundation (Grant No. 200020\_200465). P.N. acknowledges the support of the ETH Zurich Postdoctoral Fellowship Program 19-2 FEL-61.
\end{acknowledgments}

\nocite{yang20}
\bibliographystyle{unsrt}
\bibliography{apssamp}% Produces the bibliography via BibTeX.

\end{document}